\documentclass[aps,pre,amsmath,amsfonts,amssymb,superscriptaddress,preprint]{revtex4}
\usepackage{float}
\usepackage{mathtools}
\usepackage{calrsfs}
\usepackage{times}
\usepackage{latexsym}
\usepackage{graphicx}
\usepackage{amsmath,amssymb}
\usepackage{verbatim,times,bbm}
\usepackage{placeins}
\usepackage{color}
\usepackage[english]{babel}
\usepackage[T1]{fontenc}
\usepackage{graphicx}
\usepackage{amsmath, amsthm, amssymb}
\usepackage{epstopdf}
\usepackage{subfigure}
\usepackage{float}
\usepackage{appendix}
\usepackage{pdfpages}
\usepackage{placeins}
\usepackage{bm}
\usepackage{tikz}
\usetikzlibrary{positioning,shapes}

\newcommand{\affA}{Aix Marseille Univ, Université de Toulon, CNRS, CPT, Marseille, France}
\newcommand{\affB}{CNRS Centre de Physique Th\'eorique UMR7332,
13288 Marseille, France}
\newcommand{\affC}{School of Science and Technology, University of Camerino, I-62032 Camerino, Italy}
\newcommand{\affD}{INFN Sezione di Perugia, I-06123 Perugia, Italy}
\newcommand{\affE}{ QSTAR and INO-CNR, largo Enrico Fermi 2,
             I-50125 Firenze, Italy}

\begin{document}
\title{Transition between random and periodic electron currents on a DNA chain}\label{sec-intro}
 
\author{Elham Faraji}
\email{elham.faraji@unicam.it}
\affiliation{\affC}\affiliation{\affA}\affiliation{\affB}\affiliation{\affD}

\author{Roberto Franzosi}
\email{roberto.franzosi@ino.it}
\affiliation{\affE}

\author{Stefano Mancini}
\email{stefano.mancini@unicam.it}
\affiliation{\affC}\affiliation{\affD} 
 
\author{Marco Pettini}
\email{pettini@cpt.univ-mrs.fr}
\affiliation{\affA}\affiliation{\affB}

\date{\today}

\begin{abstract}
By resorting to a model inspired to the standard Davydov and Holstein-Fr\"ohlich models, 
in the present paper we study the motion of an electron along a chain of heavy particles 
modelling a sequence of nucleotides proper to a DNA fragment. Starting with a model 
Hamiltonian written in second quantization, we use the Time Dependent Variational Principle 
to work out the dynamical equations of the system. It is found that under the action of an 
external source of energy transferred to the electron, and according to the excitation site,  
the electron current can display either a broad frequency spectrum or a sharply peaked 
frequency spectrum. This sequence-dependent charge transfer phenomenology is 
suggestive of a potentially rich variety of electrodynamic interactions of DNA molecules under the 
action of electron excitation. This could imply the activation of interactions between DNA and 
transcription factors, or between DNA and external electromagnetic fields.
\end{abstract}
\maketitle
\section{Introduction}
A central question concerning the brain/mind activity and molecular dynamics in biological systems is whether in the warm, wet and noisy environment of living matter there are some processes for which quantum phenomena can play a relevant role. Some biological phenomena have been already ascribed to quantum effects, like in photosynthesis, bird orientation, light response of opsins. In the search for relevant quantum effects in living matter, electrons are certainly good candidates as is confirmed by the studies on the $\pi$-stacks of aromatic rings in proteins and in neuronal microtubules. Within this broad field of investigation, we focus on the behaviour of electrons along DNA strands. 

The possibility of charge transfer/transport along biomolecules, and notably along DNA, can have relevant biological consequences and has thus motivated several studies \cite{B0,B1,B2,B3,B4,B5,B6}. 

Following the quotations given in Ref. \cite{B6}, electron transport on DNA can be involved in the action of DNA damage response enzymes, transcription factors, or polymerase co-factors, which are relevant processes  in the cell \cite{B6,B7}. For example, evidence has been given  \cite{B6,B8}  that a DNA base excision repair enzyme enters the DNA repair process \cite{B9,B10} through an electron transfer mechanism. 
Interestingly, electron transfer through damaged regions of DNA is markedly different with respect to electron transfer through healthy regions of DNA \cite{B9}. 
 
The aim of the present work is to investigate the spectral properties of an electron current generated along a segment of DNA. The reason for this study is twofold, on the one side to investigate how DNA can respond to external electromagnetic excitations, on the other side in perspective of a further investigation about the possible activation of selective attractive forces between selected sites of a DNA sequence and transcription factors.  

Within a quantum mechanical framework, an electron transfer along some - essentially one dimensional - substrate is described by means of the probability current of the electron wave function $\psi (t)$ which, in general, reads as
\begin{eqnarray}
\vec{j}(\vec{x},t)=\frac{e \hbar}{2m_{e}i}\Big(\psi^{*}{\vec{\nabla}} \psi-\psi {\vec\nabla} \psi^{*}\Big)
\end{eqnarray}
where $e$ and $m_e$ are the electron charge and mass, respectively, and $\hbar$ is the Planck's constant; then, according to the D'Alambert equation $\partial^\mu\partial_\mu \vec{A}(t)=\mu_{0}e \vec{j}(\vec{x},t)$, where $\mu_{0}$ is the vacuum permeability, this current can generate an electromagnetic field of components 
\begin{eqnarray}
\vec{E}(t)=-\frac{\partial \vec{A}(t)}{\partial t}\nonumber\\
\vec{B}(t)=\vec{\nabla} \times \vec{A}(t)\ .
 \end{eqnarray} 
Then, the effect of the currents flowing along two macromolecules (1 and 2) can be that of generating an intermolecular force of electrodynamic kind given by the formula
\begin{equation}
\vec{F}_{12}(t) = -\frac{i_1(t) i_2(t)}{c^2} \oint\oint \frac{(d\vec{l}_1\cdot d\vec{l}_2)\vec{x}_{12}}{\vert\vec{x}_{12}\vert^3}
\label{force}
\end{equation} 
where the double integral is a geometric form factor, and the currents $i_{1,2}(t)$ are given by the mean values
\begin{equation}
i_{1,2}(t) = \frac{1}{{l}_{1,2}} \int_0^{{l}_{1,2}} \vec{j}_{1,2}(\vec{x},t)\ d\vec{x} 
\end{equation} 
where ${l}_{1,2}$ stand for the linearized lengths of molecules 1 and 2, respectively.

According to the spectral properties of these currents, and in particular in case of presence of co-resonances  in their cross frequency spectrum  
$\tilde{i}_1^*(\omega) \tilde{i}_2(\omega)$, the interaction force in Eq. (\ref{force}) could attain a strength of possible relevance in a biological context.
Thus the present work aims at contributing an ongoing research field concerning the possible activation of intermolecular electrodynamic (resonant) forces
\cite{preto,pre2,pre3}.

The present work makes a first step in this direction by providing an investigation of the spectral properties of electron currents along DNA fragments.

\section{Definition of the model and its solution}
\color{black}
There are several possible theoretical frameworks to model exciton propagation on a lattice, and the electron propagation on a lattice belongs to this class of
phenomena. A paradigmatic model is the Haken-Strobl one \cite{haken} which describes  different regimes of  exciton-phonon interaction leading to the exciton damping through the scattering by lattice vibrations and thus describing the exciton motion by some kind of hopping process. 

In the quantum biology literature, for example, the electronic energy transport to describe the coherent conveyance of electronic energy
across chromophores of protein networks - via electrodynamic coupling of their transition electric dipole moments - is described by a tight-binding Hamiltonian for an interacting $N$-body system in the presence of a single excitation \cite{kurian2}.

\color{black}
In the present work, in order to describe  electronic motions along a DNA fragment, and in perspective the related electrodynamic interactions, we resort to a model partly borrowed from the standard Davydov and Holstein-Fr\"ohlich models that have been originally introduced to account for electron-phonon interaction 
\color{black} by explicitly describing also the dynamics of the underlying molecular lattice that, at room temperature, turns out to be essentially classical. \color{black}
Thus, to model the electrons moving along a given DNA sequence, the following Hamiltonian operator is assumed  \cite{standard,froehpolaron,holstein} 
   \begin{eqnarray}\label{H}
   \hat{H} =\hat{H}_{el}+\hat{H}_{ph}+\hat{H}_{int},
   \end{eqnarray}
where the electronic and phononic parts of the Hamiltonian are given by 
   \begin{equation}\label{Hex}
   \hat{H}_{el}=\sum_{n=1}^{N}
   \Big[ E_{0}\hat{B}_{n}^{\dag} \hat{B}_{n}+\epsilon \langle\hat{B}_{n}^{\dag } \hat{B}_{n}\rangle \hat{B}_{n}^{\dag } \hat{B}_{n}+J_n(\hat{B}_{n}^{\dag} \hat{B}_{n+1}+\hat{B}_{n}^{\dag} \hat{B}_{n-1})\Big],
      \end{equation}
 \begin{equation}\label{Hph}
 \hat{H}_{ph}=\frac{1}{2}\sum_{n=1}^{N}\Big[\frac{\hat{p}_{n}^2}{M_n}+\Omega_{n}(\hat{u}_{n+1}-\hat{u}_{n})^2+{1\over 2}\mu (\hat{u}_{n+1}-\hat{u}_{n})^{4}\Big],
   \end{equation} 
and the electron-phonon Hamiltonian reads as
 \begin{equation}\label{Hin}
 \hat{H}_{int}=\sum_{n=1}^{N}\chi_{n}(\hat{u}_{n+1}-\hat{u}_{n})\hat{B}_{n}^{\dag}\hat{B}_{n}.
   \end{equation} 
As is shown in the following, the introduction of the site-dependent electron-phonon coupling constant $\chi_{n}$ brings about a definitely richer phenomenology with respect to the site-independent case \cite{scirep}.
Here, we have considered only a longitudinal sequence of nucleotides where $\hat{B}_n$ ($\hat{B}^{\dag}_n$) is the electronic annihilation (creation) operator relative to the lattice sites $n=1,2,...,N$. The term $E_{0}\hat{B}_{n}^{\dag} \hat{B}_{n}$ accounts for the initial "bare" electron energy distributed on several lattice sites according to initial shape of the electron wavefunction. The new constant $\epsilon$ is the nonlinear electron-electron coupling energy due to the interaction of the moving electron along the DNA molecule with the electrons of the substrate of nucleotides and accounts for the Coulomb repulsion between the traveling electron and the charges localized on the nucleotides. The site-dependent parameter $J_n$ determines the strength of the nearest neighbour coupling energies of the electron tunnelling across two neighbouring nucleotides.
\\
The vibronic part takes into account the longitudinal displacements of the nucleotides from their equilibrium positions, and each nucleotide  at the $n$-th site along a  DNA segment is described by the momentum and position operators, $\hat{p}_{n}$ and $\hat{u}_{n}$, by the mass $M_n$ and by the site-dependent  
$\Omega_n$, the spring parameter of two neighbouring nucleotides. The parameter $\mu$ is the coupling constant  of the nonlinear quartic term entailing the phonon-phonon interaction, of course absent in the harmonic approximation. The quartic term is introduced as a  first (stable) term - beyond the  harmonic approximation - in the power-law expansion  around the minimum of typically nonlinear interparticle interaction potential (e.g. as is the case of Van der Waals potentials). The parameter $\chi_{n}$ is the site-dependent coupling parameter of the electron-lattice interaction.\\
Here we show how the quantum equations of motion for the Hamiltonian \eqref{H} can be derived using the time dependent variational principle (TDVP) in quantum mechanics. First, we work with the second of Davydov’s ansatz state vectors by assuming the following factorization
  \begin{equation}\label{psi}
|\psi(t)\rangle=|\Psi(t)\rangle|\Phi(t)\rangle;\;\;\;\;\;\     |\Psi(t)\rangle =\sum_{n}C_{n}(t)\hat{B}_{n}^{\dag}|0\rangle_{el};\;\;\;\;\;\  |\Phi(t)\rangle=e^{-{i \over \hbar}\sum[\beta_{n}(t)\hat{p}_{n}-\pi_{n}(t)\hat{u}_{n}]}|0\rangle_{ph}
   \end{equation} 
normalized to unity by the condition $\langle\psi(t)|\psi(t)\rangle=\sum_{n}|C_{n}(t)|^2=1$, where $|C_{n}(t)|^2$ is the probability for finding the electron at the $n$-th site. The state vector $|\Psi(t)\rangle$ describes an electron given a single quantum excitation propagating along the DNA sequence of $N$ nucleotides and $|\Phi(t)\rangle$ describes the  wave function from which the expectation values of $\hat{u}_{n}$ and $\hat{p}_{n}$ are obtained as
           \begin{eqnarray}\label{22}
\langle\Phi|\hat{u}_{n}|\Phi\rangle &=&\beta_{n}(t),\nonumber\\
\langle\Phi|\hat{p}_{n}|\Phi\rangle &=&\pi_{n}(t).
   \end{eqnarray} 
 According to TDVP, which is equivalent to the least action principle, we introduce a new wave function $|\phi(t)\rangle$ -  in terms of  $|\psi(t)\rangle$ in Eq. (\ref{psi}) - by defining a time-dependent phase factor $(S(t)\in \mathbb{R})$ so that 
 \begin{eqnarray}\label{psitdvp}
|\phi(t)\rangle=e^{iS(t)/\hbar}|\psi(t)\rangle,
   \end{eqnarray} 
which implies the normalization $\langle \phi(t)|\phi(t)\rangle=1$. The wave function  $|\phi(t)\rangle$ satisfies the weaker form for the Schr\"{o}dinger equation, $i\hbar\langle\phi(t) | \partial_t|\phi(t)\rangle=\langle\phi(t) |\hat{H}|\phi(t)\rangle$, giving
           \begin{eqnarray}
-\dot{S}(t)+i\hbar\langle\psi(t) |\partial_t|\psi(t)\rangle=\langle\psi(t) |\hat{H}|\psi(t)\rangle.
   \end{eqnarray} 
Now, TDVP reads as
           \begin{eqnarray}\label{stationary}
\delta S(t)=0\;\;\;\;\;\;\; \text{with}  \;\;\;\;\;\;\; S(t)= \int_{0}^{t} L(t') dt'
   \end{eqnarray} 
where $L(t)$ is the Lagrangian associated to the system
\begin{eqnarray}\label{LL}
L(t)=i\hbar \langle\psi(t) |\partial_t|\psi(t)\rangle-\langle\psi(t) |\hat{H}|\psi(t)\rangle\ ,
   \end{eqnarray} 
hence and with equation  (\ref{psi}) we get 
    \begin{eqnarray}
L=\sum_{n}\left\{i\hbar \dot{C}_{n}(t)C_{n}^{*}(t)+{1\over2}\Big(\pi_{n}(t)\dot{\beta}_{n}(t)-\dot{\pi}_{n}(t)\beta_{n}(t)\Big)-H(C_{n},C_{n}^*,\beta_{n},\pi_{n})\right\},
   \end{eqnarray}
where 
       \begin{eqnarray}
H(C_{n},C_{n}^*,\beta_{n},\pi_{n})= \langle\psi(t) |\hat{H}|\psi(t)\rangle.
   \end{eqnarray}
By requiring the fulfilment of the stationary action condition of Eq. \eqref{stationary}, we get 
\begin{eqnarray}
\delta S(t)&=\sum_{n}\Big\{ i\hbar\Big(-\dot{C}_{n}^{*}(t)\delta C_{n}(t)+\dot{C}_{n}(t)\delta C_{n}^{*}(t)\Big)+\dot{\beta}_{n}(t)\delta \pi_{n}(t)-\dot{\pi}_{n}(t)\delta \beta_{n}(t)\nonumber\\
&-(\partial_{C_{n}}H)\delta C_{n}-(\partial_{C_{n}^{*}}H)\delta C_{n}^{*}-(\partial _{\beta_{n}}H)\delta \beta_{n}-(\partial_{\pi_{n}}H)\delta \pi_{n}\Big\}=0,
   \end{eqnarray}
    and arrive at the dynamical equations 
           \begin{eqnarray}\label{hc}
i\hbar\dot{C}_{n}&=&\partial_{C^*_{n}} H\nonumber\\
\dot{\beta}_{n}&=&\partial _{\pi_{n}} H\nonumber\\
\dot{\pi}_{n}&=&-\partial _{\beta_{n}} H \ .
   \end{eqnarray}
The expectation value of the Hamiltonian is
      \begin{eqnarray}\label{Hnew}
\langle \psi|\hat{H}|\psi \rangle&=\sum_{n}\Big[E_{0}|C_{n}|^{2}+\epsilon|C_{n}|^{4}+ J_n(C_{n}^{*}C_{n+1}+C_{n+1}^{*}C_{n})\nonumber\\
&+ \frac{1}{ 2}\Big(\frac{1}{M_n}\pi_{n}^2+\Omega_{n} (\beta_{n+1}-\beta_{n})^2+{1\over 2}\mu (\beta_{n+1}-\beta_{n})^4\Big)\nonumber\\
&+\chi_n(\beta_{n+1}-\beta_{n})|C_{n}|^2\Big].
   \end{eqnarray}
Thus, from (\ref{hc}) and (\ref{Hnew}), we find the following explicit form of the equations of motion 
 \begin{eqnarray}\label{dynamicaleqs}
  i\hbar\dot{C}_{n}&=&\Big(E_{0}+2\epsilon|C_{n}|^{2}+\chi_n(\beta_{n+1}-\beta_{n})\Big)C_{n}
 +J_{n} C_{n+1}+J_{n-1}C_{n-1},\nonumber\\
 M_{n}\ddot{\beta}_{n}&=&\Omega_{n}\beta_{n+1}+\Omega_{n-1}\beta_{n-1}-\Omega_{n-1}\beta_{n}-\Omega_{n}\beta_{n}+\chi_n |C_{n}|^2-\chi_{n-1} |C_{n-1}|^2\nonumber\\
&+&\mu \Big((\beta_{n+1}-\beta_{n})^{3}-(\beta_{n}-\beta_{n-1})^{3}\Big).
    \end{eqnarray}
  \section{Physical parameters}  
  \begin{center}
\begin{table}[htbp]\label{energies}
\centering
\begin{tabular}{|c|c|c||c|c|c|}
\hline
          \textbf{Nucleotide} &\textbf{EIIP Ry}&\textbf{EIIP eV}& \textbf{Nucleotide} &\textbf{EIIP Ry}&\textbf{EIIP eV}
        \\  
        \hline
       A&0.1260& 1.7143&     T&0.1335&  1.8164\\
     G&0.0806&  1.0966&      C&0.1340&  1.8232   \\    
   \hline              
\end{tabular}
\caption{Electron-Ion interaction potential (EIIP) values for nucleotides adenine (A), thymine (T), guanine (G), and cytosine (C). From Ref.\cite{Cosic}.}
\end{table}
\end{center}
\vspace{ -1.5cm}
 {{In order to choose meaningful physical coupling parameters of the Hamiltonian, we borrow from Ref. \cite{Cosic,pseudopot} the values of the interaction energy between an electron and each of all the 4 nucleotides (reported in Table I). We assume that the moving electron - of initial energy $E_0$ - moves along the sequence of nucleotides constituting a given segment of DNA  by tunnelling across  square potential barriers of variable height and of width $a = 3.4 \mathring{A}$, the average distance between two nearest neighbouring nucleotides \cite{standard}. We introduce the transmission coefficients $ T_{n,n+1}$ from the probability $P(n\to n\pm 1)$ of tunnelling from one potential well to the nearest one in order to set the electron hopping terms $J_{n}$ in (\ref{Hex}) 
 \begin{equation}
 T_{n,n+1}= \left[ 1 + \frac{ E^{2}_{n+1} \sinh^2(\beta_{n+ 1} a)}{4 E_0(E_{n+ 1}  - E_0)}\right]^{-1}\ ,
 \end{equation}
where $\beta_{n+ 1} =[2 m_e(E_{n+ 1} - E_0)/\hbar^2]^{1/2}$, $m_e$ is the mass of electron and $E_{n+ 1}$ are the interaction energies between the moving electron and the local nucleotide. Then we assume 
\begin{eqnarray}\label{Jn}
J_{n} = E_{0} T_{n,n+ 1}.
\end{eqnarray}
For the interaction Hamiltonian (\ref{Hin}) we can roughly estimate the electron-phonon coupling parameter $\chi_{n}$ as
\begin{eqnarray}\label{chin}
\chi_{n}=d E/d x =\frac{E_{n+1}-E_{n}}{a}.
\end{eqnarray}
Numerical simulations are performed by adopting dimensionless physical parameters in the dimensionless expressions of the Hamitonian (\ref{Hnew}) and of the dynamical equations (\ref{dynamicaleqs}). These are  found by rescaling time and lengths as $t=\omega^{-1} \tau$ and $\beta_{n}=L b_{n}$, respectively, where $L=\sqrt{\hbar \omega^{-1}M_{n}^{-1}}$. The outcomes are 
         \begin{align}\label{Hsimpli}
\langle \psi|\hat{H}|\psi \rangle&=\sum_{n=1}^N\Big[E'|C_{n}|^{2}+\epsilon'|C_{n}|^{4}+ J_{n}'(C_{n}^{*}C_{n+1}+C_{n+1}^{*}C_{n})\nonumber\\
&+ {1\over 2}\Big(\dot{b}^{2}_{n}+ \Omega_{n}'(b_{n+1}-b_{n})^2+{1\over 2}\mu'(b_{n+1}-b_{n})^4\Big)\nonumber\\
&+\chi_{n}'(b_{n+1}-b_{n})|C_{n}|^2\Big],
   \end{align}
and 
\begin{eqnarray}\label{dimdynamics}
  i\frac{d{C}_{n}}{d\tau}&=&\Big(E'+2\epsilon'|C_{n}|^{2}+\chi_{n}'(b_{n+1}-b_{n})\Big)C_{n}
 +J_{n}'C_{n+1}+J_{n-1}'C_{n-1},\nonumber\\
\frac{d^{2}{b}_{n}}{d\tau^2}&=&\Omega_{n}'b_{n+1}+\Omega_{n-1}' b_{n-1}-\Omega_{n-1}'b_{n}-\Omega_{n}'b_{n}+\chi_{n}'|C_{n}|^2-\chi_{n-1}'|C_{n-1}|^2\nonumber\\
&+&\mu'\Big[(b_{n+1}-b_{n})^3-(b_{n}-b_{n-1})^3\Big],
   \end{eqnarray}
where
   \begin{eqnarray}\label{dimparameter}
   E'&=&{E_{0} \over \hbar \omega};\;\;\;\;\;\;\; \epsilon'={\epsilon \over \hbar \omega};\;\;\;\;\;\;\; J_{n}'={J_{n}\over \hbar \omega};\;\;\;\;\;\;\; \nonumber\\
    \chi_{n}'&=&{\chi_{n} \over \sqrt{\hbar M_{n}\omega^{3}}};\;\;\;\;\; \Omega_{n}'={\Omega_{n}\over M_{n} \omega^2};\;\;\;\;\;\;\mu'={\mu \hbar \over M_{n}^{2}\omega^{3}}\ .
   \end{eqnarray}
   In our simulations we resort to the known sound speed $V=a(\Omega_{n}/M_{n})^{1/2}$ on DNA; we borrow the value $V=1.69$ km/s from \cite{Hakim}, and
  take it as an average constant (neglecting small local variations due to the different masses of the nucleotides). Thus we obtain the constant dimensionless parameter $\Omega'=V^{2}/a^{2}\omega^2$ from (\ref{dimparameter}). In re-writing the dynamical equations we introduce the variables
           \begin{eqnarray}
q_{n}={C_{n}+C_{n}^{*}\over \sqrt{2}}, \qquad p_{n}={C_{n}-C_{n}^{*}\over i\sqrt{2}}\ ,
   \end{eqnarray}
so that Eqs. (\ref{dimdynamics}) become
\begin{align}
           \dot{q}_{n}&=\Big[E'+{\epsilon' \over 2}(q_{n}^{2}+p_{n}^{2})+\chi'_n(b_{n+1}-b_{n})\Big]p_{n}+J_{n}'p_{n+1}+J_{n-1}'p_{n-1},\label{differential28}\\
\dot{p}_{n}&=-\Big[E'+{\epsilon' \over 2}(q_{n}^{2}+p_{n}^{2})+\chi'_n(b_{n+1}-b_{n})\Big]q_{n}+J_{n}'q_{n+1}+J_{n-1}'q_{n-1}\Big], \label{differential29}\\
\ddot{b}_n&=\Omega'(b_{n+1}+b_{n-1}-2b_{n})+ {1\over 2}  \Big(
\chi_{n}'(q_{n}^{2}+p_{n}^{2})-\chi_{n-1}'(q_{n-1}^{2}+p_{n-1}^{2})\Big)\nonumber\\
&+\mu'\Big[(b_{n+1}-b_{n})^3-(b_{n}-b_{n-1})^3\Big]\label{differentialqp}\nonumber\\
&=\mathcal{B}_n[\textbf{b}(t), \textbf{q}(t), \textbf{p}(t)]\ .
   \end{align}
where the equation for $\ddot{b}_n$ can be also written as 
\begin{eqnarray}\label{bienne}
\dot{b}_n&=& \pi_n\nonumber\\
\dot{\pi}_n&=& \mathcal{B}_n[\textbf{b}(t), \textbf{q}(t), \textbf{p}(t)]\ .
\end{eqnarray}
Finally, by combining a leap-frog scheme for \eqref{bienne} and a finite differences scheme for $\dot{q}_{n}$ and $\dot{p}_{n}$,  equations \eqref{differential28}-\eqref{differentialqp} are rewritten in a form suitable for numerical solution, that is 
   \begin{eqnarray}\label{eqdinamiche}
q_{n}(t+\Delta t)&=& q_{n}(t)+\Delta t\ \mathcal{Q}_n[\textbf{b}(t), \textbf{q}(t), \textbf{p}(t)],\nonumber\\
p_{n}(t+\Delta t)&=& p_{n}(t)+\Delta t\ \mathcal{P}_n[\textbf{b}(t), \textbf{q}(t), \textbf{p}(t),\nonumber\\
b_{n}(t+\Delta t)&=& b_{n}(t)+\Delta t\ \pi_n(t),\nonumber\\
\pi_{n}(t+\Delta t)&=& \pi_{n}(t)+\Delta t\ \mathcal{B}_n[\textbf{b}(t+\Delta t), \textbf{q}(t+\Delta t), \textbf{p}(t+\Delta t)] ,
   \end{eqnarray}
   where $\mathcal{Q}_n[\textbf{b}(t), \textbf{q}(t), \textbf{p}(t)]$ and $\mathcal{P}_n[\textbf{b}(t), \textbf{q}(t), \textbf{p}(t)$ are the r.h.s. of Eqs. \eqref{differential28} and \eqref{differential29}, respectively. The integration scheme for $b_{n}(t)$ and $\pi_n(t)$ is a symplectic one, meaning that all the Poincar\'e invariants 
of the associated Hamiltonian flow are conserved, among these invariants there is energy. 
The simple leap-frog scheme cannot be applied to the equations for $\dot{q}_{n}(t)$ and $\dot{p}_{n}(t)$ because the r.h.s. of the equations for $\dot{q}_{n}(t)$ explicitly depend on $q_{n}(t)$ and $b_{n}(t)$, therefore  the first two equations in \eqref{eqdinamiche} are integrated with an Euler predictor-corrector to give
    \begin{eqnarray}\label{eqdinamicheBis}
q^{(0)}_{n}(t+\Delta t)&=& q_{n}(t)+\Delta t\ \mathcal{Q}_n[\textbf{b}(t), \textbf{q}(t), \textbf{p}(t)],\nonumber\\
p^{(0)}_{n}(t+\Delta t)&=& p_{n}(t)+\Delta t\ \mathcal{P}_n[\textbf{b}(t), \textbf{q}(t), \textbf{p}(t)],\nonumber\\
q^{(1)}_{n}(t+\Delta t)&=& q_{n}(t)+\frac{\Delta t}{2}\left\{ \mathcal{Q}_n[\textbf{b}(t), \textbf{q}(t), \textbf{p}(t)] +  \mathcal{Q}_n[\textbf{b}(t), \textbf{q}^{(0)}(t+\Delta t), \textbf{p}^{(0)}(t+\Delta t)]\right\},\nonumber\\
p^{(1)}_{n}(t+\Delta t)&=& p_{n}(t)+\frac{\Delta t}{2}\left\{\mathcal{P}_n[\textbf{b}(t), \textbf{q}(t), \textbf{p}(t)] +  \mathcal{P}_n[\textbf{b}(t), \textbf{q}^{(0)}(t+\Delta t)), \textbf{p}^{(0)}(t+\Delta t)]  \right\},\nonumber\\
b_{n}(t+\Delta t)&=& b_{n}(t)+\Delta t\ \pi_n(t),\nonumber\\
\pi_{n}(t+\Delta t)&=& \pi_{n}(t)+\Delta t\ \mathcal{B}_n[\textbf{b}(t+\Delta t), \textbf{q}^{(1)}(t+\Delta t), \textbf{p}^{(1)}(t+\Delta t)] ,
   \end{eqnarray}
so that, thanks to fact that half of the set of the dynamical equations \eqref{eqdinamiche} is integrated by means of a symplectic algorithm, and half of the equations are integrated by means of the Euler predictor-corrector, it turns out that by adopting sufficiently small integration time steps $\Delta t$ the total energy is very well conserved without any drift, just with zero-mean fluctuations around a given  value fixed by the initial conditions.
 About initial conditions, independently of the specific physical excitation mechanism, we assume an electron wavefunction described by the amplitudes $C_{n}(t=0)$ centered at the excitation site $n=n_{0} $ and distributed at time $t=0$ \cite{standard} as 
         \begin{eqnarray}\label{exc}
C_{n}(t=0)&=& {1\over \sqrt{8 \sigma_{0}}}{\rm sech}\Big({n-n_{0}\over 4\sigma_{0} }\Big)
    \end{eqnarray}\\
where $\sigma_{0}$ specifies the width of the function. Periodic boundary conditions have been used.

About the initial conditions of the phonon part of the system, we assume that the components of the DNA fragment under consideration  are initially thermalized at the room temperature $T =310 K$. At thermal equilibrium, the average kinetic and potential energy per degree of freedom are equal, therefore at  $t=0$ the displacements and the associated velocities have been initialized with random values of zero mean and according to the following prescription 
   \begin{equation}\label{thermequil}
\langle\vert b_{n}(0)\vert\rangle_n=\sqrt{\frac{k_{B}T}{\hbar \omega \Omega'}} ;\;\;\;\;\;\;\;\;\;\;\;\;\;\;\;\;\; \langle\vert \dot{b}_{n}(0)\vert\rangle_n=\sqrt{\frac{k_{B}T}{\hbar \omega}}.
    \end{equation}
expressed in dimensionless form. Periodic boundary conditions have been used also for the phonon part of the system.
 \section{Numerical results}
In numerical simulations, we have adopted an integration time step  $\Delta t=5 \times 10^{-6}$ (dimensionless units) entailing a very good energy conservation with typical fluctuations  of relative amplitude $\Delta E/E\simeq 10^{-6}$. 

In what follows, we report the results  obtained for two sequences of $N=66$ and $N =398$ nucleotides, respectively, for different initial electron energies $E_0$, 
and for initial excitation sites $n_0$ entering the initial wavefunction shape (\ref{exc}). 

As motivated in the Introduction, the physical quantity of main interest in what follows is the Fourier spectrum of the electron current activated on a segment of DNA. The average electron current is derived from the standard probability current $j(x_i,t)$ associated with the wave function $|\psi (t)\rangle$ in \eqref{psi}, which, in discretized version along the chain of nucleotides and thus ready for its numerical computation, reads as 
\begin{eqnarray}
{j}(x_i,t)=\frac{e \hbar}{2m_{e}i}\Big(\psi^{*}(x_i,t)\frac{\psi(x_{i+1},t)-\psi(x_{i-1},t)}{2a} -\psi(x_i,t)\frac{\psi^*(x_{i+1},t)-\psi^*(x_{i-1},t)}{2a}\Big)
\end{eqnarray}
and hence the average current
\begin{equation}
i_{av}(t) = \frac{1}{l} \int_0^l  j(x,t) dx = \frac{1}{N a}\sum_{i=1}^N j(x_i,t) a \ .
\end{equation} 

In Figures \ref{amplitude3452} - \ref{amplitude3434} the outcomes are reported of the simulations performed for a DNA  sequence coding for a subunit of the human haemoglobin molecule (HBB) consisting of $N=398$ nucleotides. Figures \ref{amplitude3452} and \ref{amplitude3440} display the behaviour of the system when $E_{0}=0.658$ eV and for $n_{0}=N/2$ and $n_{0}=N/3$, respectively.
Remarkably,  the mean electron current is alternate in what it takes positive and negative values, however, the corresponding Fourier power spectra appear very different according to the initial excitation site. In fact, for $n_{0}=N/3$ it is well evident that the spectrum $\vert\tilde{i}(\nu)\vert^2$ is peaked around the frequency 
$\nu\simeq 5$ THz, whereas for $n_{0}=N/2$ the spectrum appears much broader and noisy.

Then, by comparing the results obtained keeping the initial excitation around the site $n_{0}=N/3$ but increasing the energy to $E_0=0.79$ eV, an interesting and to some extent surprising result is found: the power spectrum of the current gets sharper and peaked around $\nu\simeq 44$ THz, a much higher frequency indeed. This is shown in Figure \ref{amplitude3441}.

On the other hand, with the initial excitation again localized around the site $n_{0}=N/2$ and with a lower initial activation energy, $E_0=0.46$ eV, Figure \ref{amplitude3434} shows that  the electron wave function quickly spreads  over the whole chain of nucleotides and the electron current power spectrum  broadens with respect to that one found for $E_0=0.658$ eV.  Of course, the parameter space of the system is very large and thus its systematic investigation is practically unfeasible. Nevertheless, the above reported results outline the existence of a richer phenomenology with respect to the excitation of just Davydov electro-solitons. 
This could be of interest in view of modelling specific processes involving electrodynamic interactions of DNA with other biomolecules or with external electromagnetic fields.

Panels (b) of the different figures show the time evolution of random initial conditions for the displacements of the underlying sequences of masses modelling nucleotides of the DNA. According to the prescriptions of Eq. (\ref{thermequil}), the random initial displacements and velocities are made at thermal equilibrium at $310^{\circ}$ K.

In Figures \ref{amplitude3423} - \ref{amplitude3432} the results are  reported of numerical simulations obtained for a shorter sequence of $N=66$ nucleotides
containing the GAATTC recognition site of the restriction endonuclease enzyme EcoRI \cite{eigen,vitiello}. These results confirm the richness of the phenomenology previously observed for the longer DNA sequence. The spatial distribution of the probability $|\Psi(x,t)|^2$ in Fig. \ref{amplitude3423} where $E_0=0.2$ eV and $n_0=N/3$ seems completely frozen in time even though some low amplitude ripple exists which entail a non-vanishing electron current; the power spectrum 
$\vert\tilde{i}(\nu)\vert^2$ shows some peaks concentrated in the frequency interval  $5-10$ THz. 
By increasing the initial excitation energy of the electron above $0.6$ eV, and keeping the excitation centered at the same site $n_0=N/3$, we can see  in Figure  \ref{amplitude3426} a quick and complete spreading of the electron wave function $|\Psi(x,t)|^2$ and, correspondingly, a broad and noisy power spectrum of the electron current. 
A further increase of the electron excitation energy at $E_0=0.79$ eV, with the initial wavefunction centered at the site $n_0=N/2$, brings about a well evident ripple of $|\Psi(x,t)|^2$ around its initial peak, as it can be seen in Figure (\ref{amplitude3421}). The associated electron current power spectrum turns out to be extremely peaked around  $48$ THz.  
Then, considering the initial electron energy at $E_0=0.9$ eV and displacing the initial excitation around the site $n_0=2N/3$ it is observed that the peak value of the electron wavefunction decreases much faster than in the previous case ($n_0=N/2$) and also the power spectrum of the current is very different from the previous case, displaying a broad and noisy pattern, as can be seen in Figure (\ref{amplitude3432}).
\color{black}
A comment about the possible biological significance of the energies adopted for the simulations is in order. The electron excitation values of $0.2$, $0.46$, and 
$0.6$ eV correspond to one, two and three  DNA phosphodiester bonds, respectively, and are the same of those considered in \cite{kurian3} where an analysis has been carried on for the same restriction enzyme DNA target sequence considered in the present work. 
\color{black}

A complementary characterization of the electron current power spectra can be performed by means of spectral entropy. This is defined \textit{\`a la} Shannon as follows
\begin{equation}
S(t)=- \sum_{m=1}^{\text{M}/2}w_m(t) \ln w_m(t);\;\;\;\;\;\;\;\; w_m(t) =\frac{ E_m(t)}{E_{T}(t)}
\end{equation}
 the weights $w_m(t) $ are normalized by 
\begin{equation}
E_{T}(t)=\sum_{m=1}^{\text{M}/2}E_m(t);\;\;\;\;\;\;\;\; E_m(t) =\nu_{m}^2 |\tilde{i}(\nu_{m})|^2
\end{equation}
where $\tilde{i}(\nu_m)$ indicates the Fourier transform of the electron current. The frequencies are $\nu_{m}={m/T}=m/(M \Delta T)$ where $T$ is the length of  time window which is Fourier analyzed, $\Delta T$ is a sampling time such that $T= M \Delta T$. As the input signals are real numbers and the Fourier spectra are
computed through DFT algorithm we ignore the mirror part of the spectra thus ignoring the second half of the FFT.
Then the spectral entropy is normalized to give
\begin{equation}
\eta(t)=\frac{S_{max}(t)-S(t)}{S_{max}(t)}
\end{equation}
so that it is $\eta=1$ when the power spectrum of the electron current is monochromatic,  and $\eta=0$ for a flat spectrum such that $S(t)=S_{max}(t)$.  Figure  \ref{entropy} shows the normalized entropy $\eta$ of the electrons versus the initial energy $E_0$.  In panel (a)  $\eta$ versus $E_0$ is reported for the longer DNA segment under different initial conditions:  for $n_0=N/2$ the normalized entropy $\eta$ takes values approximately in the interval $0.1 - 0.15$ meaning that the corresponding power spectra are broad and noisy; for initial excitation site $n_0=N/3$ it is found $\eta=0.45$ at $E_{0}=0.79$ eV and we recover what has been already displayed by Figure \ref{amplitude3441}, that is, the power spectrum is narrow in frequency; intermediate values of $\eta$ are found when the excitation site is $n_0=2N/3$ at energies $E_0=0.66, 0.79, 0.9$ eV. 

 In panel (b)  $\eta$ versus $E_0$ is reported for the shorter DNA  segment and synoptically shows the nontrivial appearance of some kind of, loosely speaking,  "resonances" in what $\eta$ displays some bumps in its $E_0$-pattern which correspond to a significant narrowing of the electron current spectra, and thus to a less noisy and more coherent behaviour of the electron current. 
 
\section{Concluding remarks}  
The novelty of the model investigated in the present work concerns the introduction of  site dependent electron-phonon coupling constants. The sequence of values of these coupling constants follows the sequence of nucleotides along a given DNA segment. Also the sequence of probabilities of electron jumping from one site to the next one depends on the specific sequence of nucleotides. In so doing a richer phenomenology is found with respect to a similar model recently investigated \cite{scirep}. Instead of observing the propagation of  a standard Davydov electro-soliton, depending on the initial excitation energy, we have observed localized periodic motions of the electrons - giving rise to a narrow frequency spectrum of the electron current - or, depending on the initial excitation site, to more extended motions associated with a broad noisy frequency spectrum. In both cases, the relevant spectral range belongs to the THz frequency domain. A qualitatively similar phenomenology has been found by tackling two different DNA molecules, a subunit of the haemoglobin molecule (HBB) and an oligonucleotide with a specific recognition site of a restriction enzyme. Our findings suggest that the activation of periodic currents on specific sites could be  at the origin of attractive forces between the DNA and a specific effector (transcription factor, enhancer, inhibitor, and so on). The prospective developments of the present work thus concern a new investigation/explanation of the physical grounds of the Resonant Recognition Model \cite{Cosic}  by looking for co-resonances in the current frequency spectra of biochemical  reaction partners. Moreover, electromagnetic signalling from electronic currents flowing along a DNA strand, and DNA response to externally applied electromagnetic fields could be further investigated in the light of the approach proposed in the present work.
\color{black}
Finally, further developments of the present model, aimed at describing co-resonance-activated intermolecular interactions, will have to take into account the role of hydration layers of the interacting bio-molecules. In fact, as put forward in \cite{preto}, hydration layers made of spatially ordered water dipole moments can deeply affect the strength of the intermolecular electrodynamic interactions and the existence and relevance of these hydration layers is experimentally proved for DNA molecules \cite{dermott,elia} and for proteins \cite{susko}.      
\color{black}

\section*{Acknowledgments}
E.F. warmly thanks the Fondazione Cassa di Risparmio di Firenze for having co-funded her PhD fellowship.
M.P. participated in this work within the framework of the project MOLINT which has received funding from the Excellence Initiative 
of Aix-Marseille University - A*Midex,  a French “Investissements d’Avenir” programme. 
R.F. acknowledges support from the QuantERA ERA-NET Co-fund 731473 (Project Q-CLOCKS), and from the italian National Group of Mathematical Physics (GNFM-INdAM).

\begin{figure}
\includegraphics[width=0.45\columnwidth]{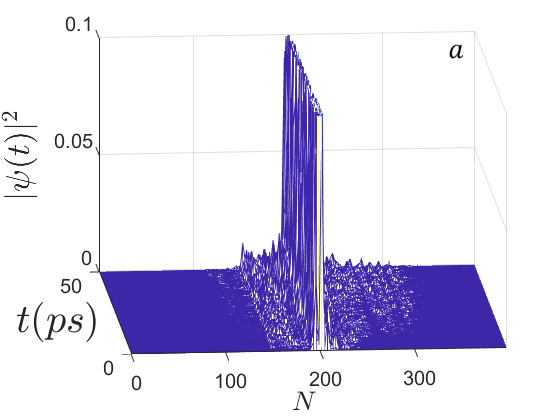}
\includegraphics[width=0.45\columnwidth]{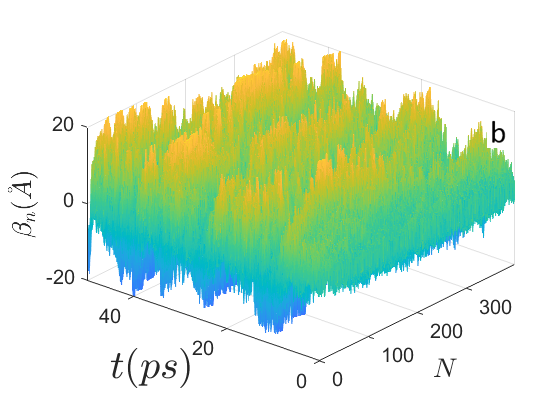}
\includegraphics[width=0.45\columnwidth]{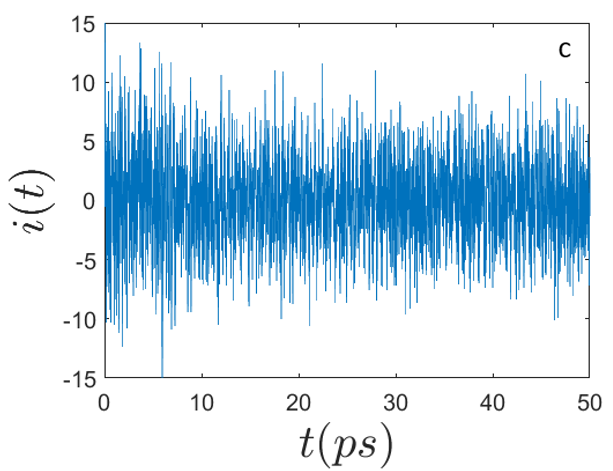}\hskip 1.5truecm
\includegraphics[width=0.45\columnwidth]{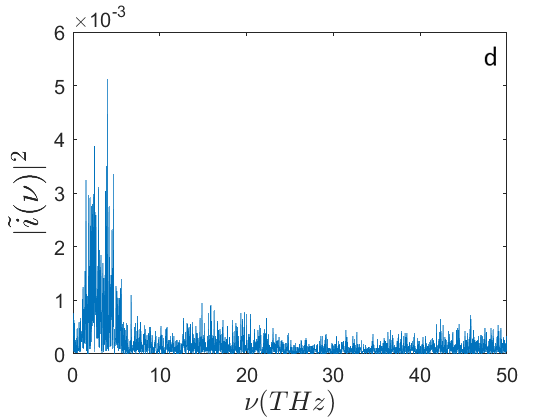}      
   \caption{Time evolution of: a) the electron probability amplitude $\vert\Psi(t)\vert^2$; b) the average displacements $\beta_{n}(\mathring{A})$; c) the electron current $i(t)$ (expressed in $\mu$biot); d) the frequency spectrum $\vert\tilde{i}(\nu)\vert^2$ along the sequence of $N=398$ nucleotides. Initial conditions: $T=310^{\circ}$K, $n_0=N/2$, $E'=100$, $\epsilon'=5$,  $\Omega'=0.25$, $\mu'=0.5$, $\sigma_{0}=0.1$, and the site-dependent parameters $J'_n$,  and $\chi'_n$ corresponding to $E_0=0.658$ eV, $\epsilon=0.0329$ eV,  $\Omega_n=V^{2}M_n/a^2$, $J_n$,  and $\chi_n$ in Eqs. (\ref{Jn}) and (\ref{chin}), respectively. Initial electron excitation peaked around the site $n_0=N/2$.}
 \label{amplitude3452}
 \end{figure}

\begin{figure}
\includegraphics[width=0.45\columnwidth]{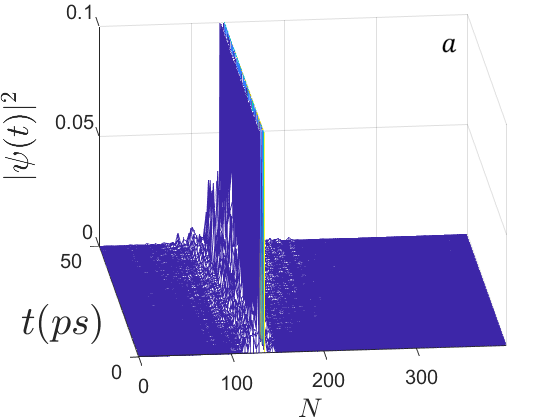}
\includegraphics[width=0.45\columnwidth]{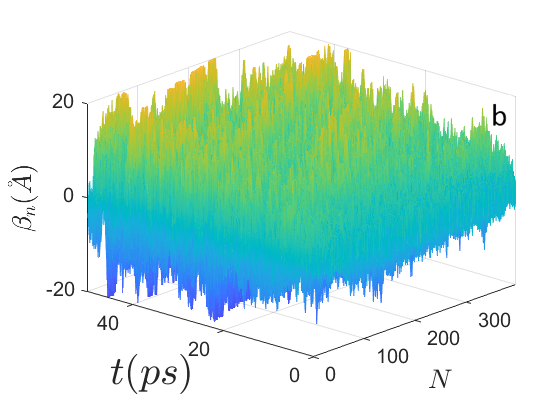}
\includegraphics[width=0.45\columnwidth]{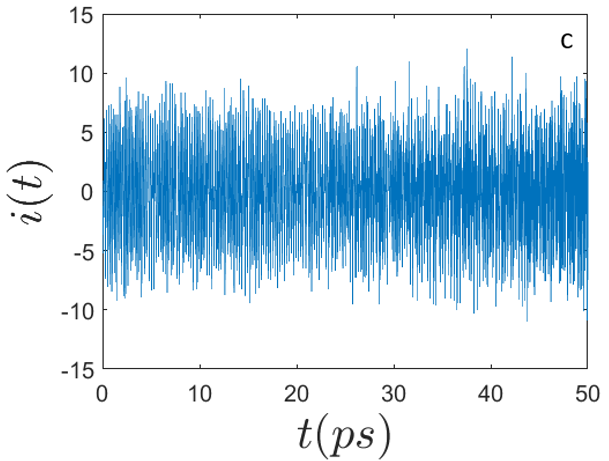}\hskip 1.5truecm
\includegraphics[width=0.45\columnwidth]{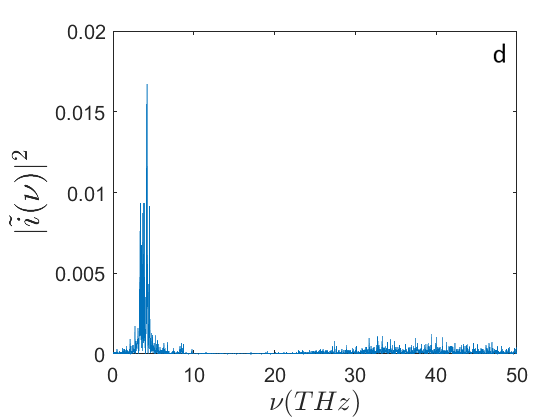}      
 \caption{Time evolution of: a) the electron probability amplitude $\vert\Psi(t)\vert^2$; b) the average displacements $\beta_{n}(\mathring{A})$; c) the electron current $i(t)$ (expressed in $\mu$biot); d) the frequency spectrum $\vert\tilde{i}(\nu)\vert^2$ along the sequence of $N=398$ nucleotides. Initial conditions and parameters are the same of Fig. \ref{amplitude3452} but the initial electron excitation is peaked around the site $n_0=N/3$.}
 \label{amplitude3440}
 \end{figure}

\begin{figure}
\includegraphics[width=0.45\columnwidth]{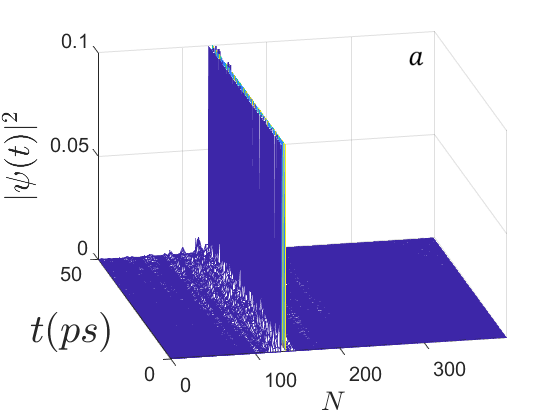}
\includegraphics[width=0.45\columnwidth]{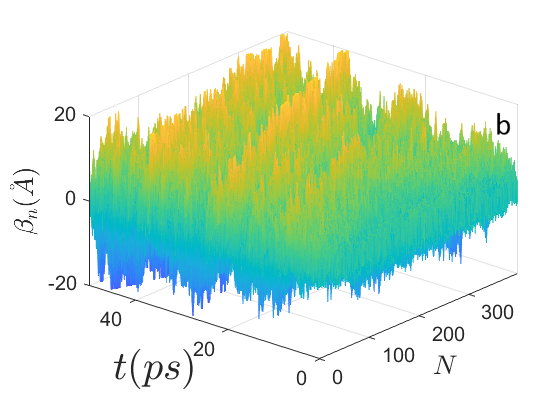}
\includegraphics[width=0.45\columnwidth]{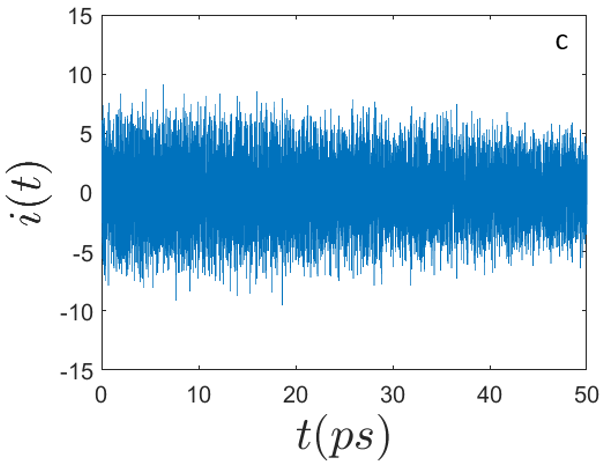}\hskip 1.5truecm
\includegraphics[width=0.45\columnwidth]{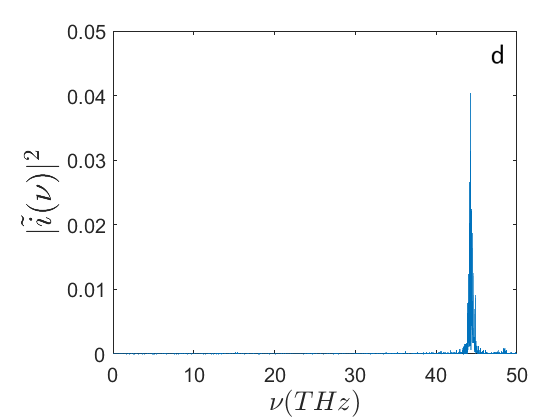}      
 \caption{Time evolution of: a) the electron probability amplitude $\vert\Psi(t)\vert^2$; b) the average displacements $\beta_{n}(\mathring{A})$; c) the electron current $i(t)$ (expressed in $\mu$biot); d) the frequency spectrum $\vert\tilde{i}(\nu)\vert^2$ along the sequence of $N=398$ nucleotides, with the initial condition  $E_0=0.79$ eV ($E'=120$) and the initial electron excitation peaked around the site $n_0=N/3$.  The other parameters are the same of Fig. \ref{amplitude3452}.}
 \label{amplitude3441}
 \end{figure}

\begin{figure}
\includegraphics[width=0.45\columnwidth]{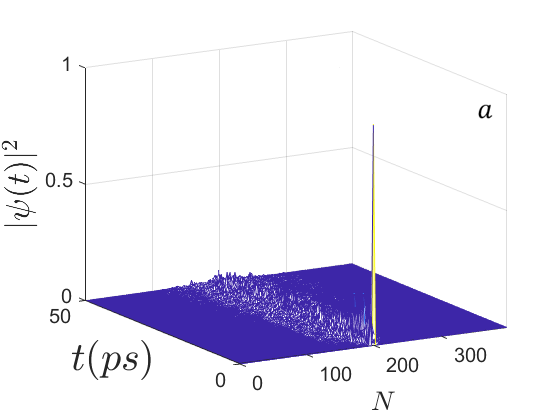}
\includegraphics[width=0.45\columnwidth]{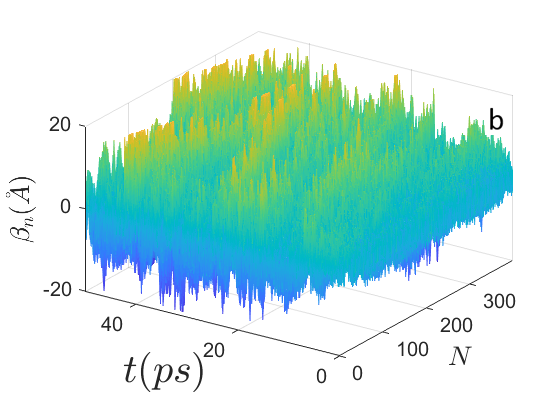}
\includegraphics[width=0.45\columnwidth]{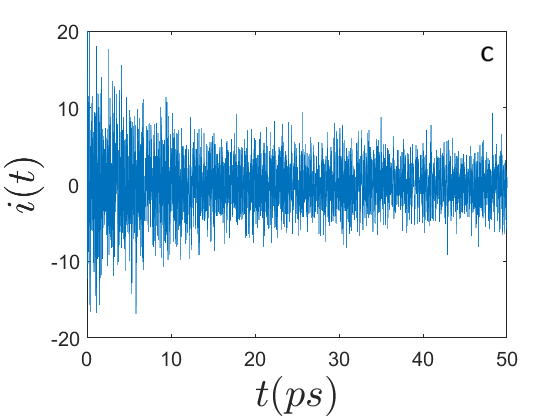}\hskip 1.5truecm
\includegraphics[width=0.45\columnwidth]{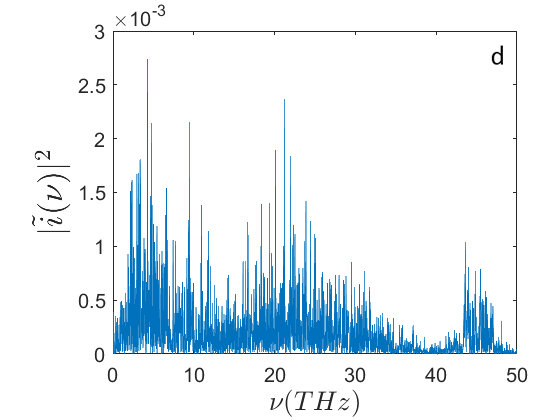}      
 \caption{Time evolution of: a) the electron probability amplitude $\vert\Psi(t)\vert^2$; b) the average displacements $\beta_{n}(\mathring{A})$; c) the electron current $i(t)$ (expressed in $\mu$biot); d) the frequency spectrum $\vert\tilde{i}(\nu)\vert^2$ along the sequence of $N=398$ nucleotides with the initial condition $E_0=0.46$ eV ($E'=70$).  The initial electron excitation is peaked around the site $n_0=N/2$. The other parameters are the same of Fig. \ref{amplitude3452}.}
 \label{amplitude3434}
 \end{figure}

\begin{figure}
\includegraphics[width=0.45\columnwidth]{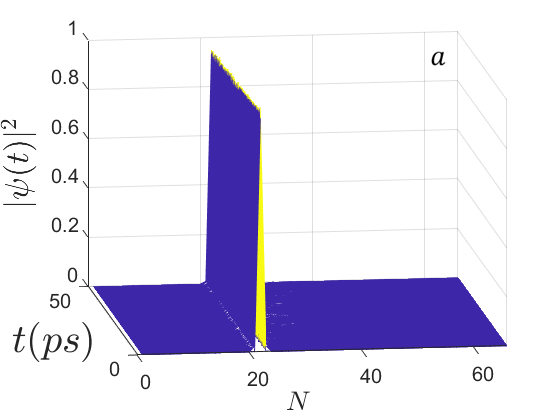}
\includegraphics[width=0.45\columnwidth]{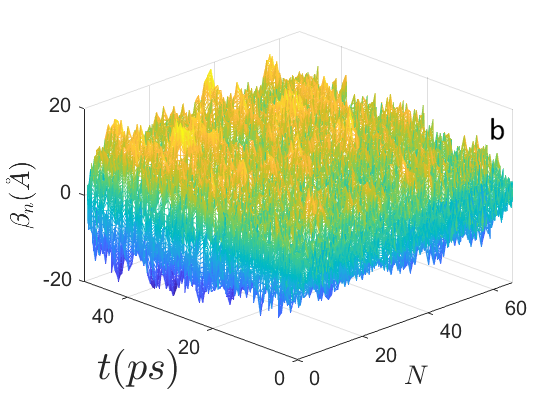}
\includegraphics[width=0.45\columnwidth]{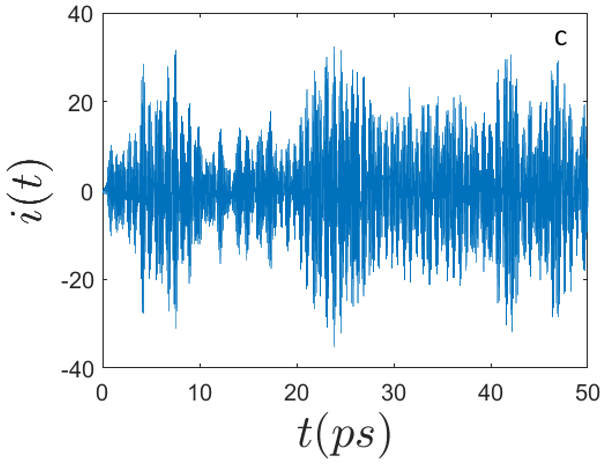}\hskip 1.5truecm
\includegraphics[width=0.45\columnwidth]{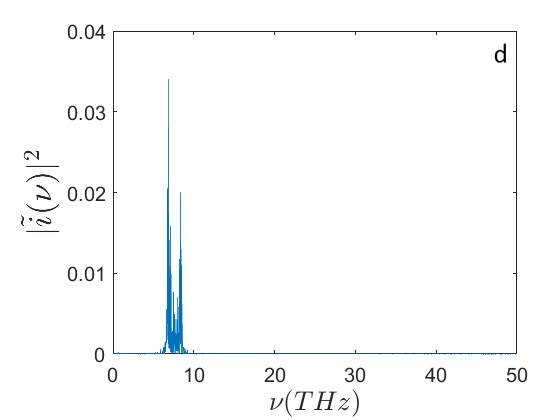}      
   \caption{Time evolution of: a) the electron probability amplitude $\vert\Psi(t)\vert^2$; b) the average displacements $\beta_{n}(\mathring{A})$; c) the electron current $i(t)$ (expressed in $\mu$biot); d) the frequency spectrum $\vert\tilde{i}(\nu)\vert^2$ along the sequence of $N=66$ nucleotides with the initial conditions $E_0=0.2$ eV ($E'=30$) and the initial electron excitation peaked around the site $n_0=N/3$. The other parameters are the same of Fig.\ref{amplitude3452}.}
 \label{amplitude3423}
 \end{figure}

\begin{figure}
\includegraphics[width=0.45\columnwidth]{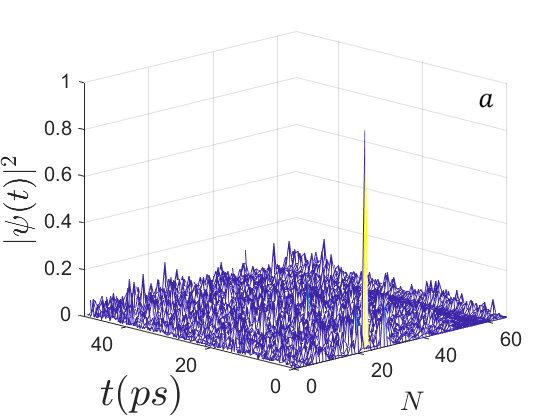}
\includegraphics[width=0.45\columnwidth]{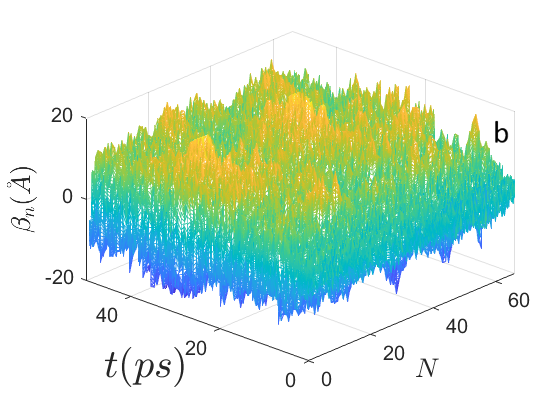}
\includegraphics[width=0.45\columnwidth]{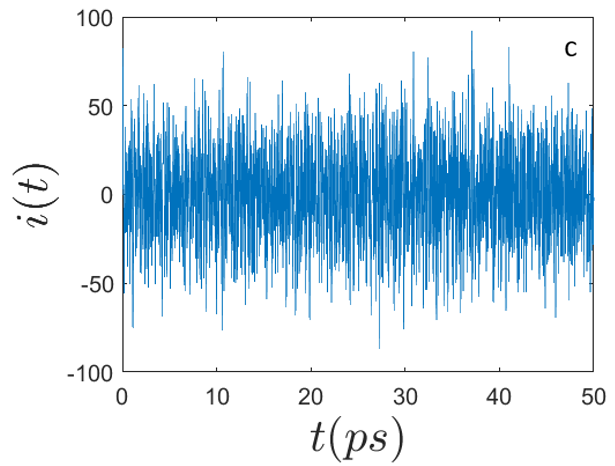}\hskip 1.5truecm
\includegraphics[width=0.45\columnwidth]{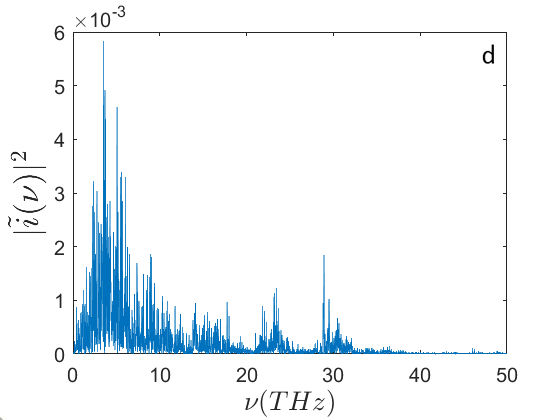}      
   \caption{Time evolution of: a) the electron probability amplitude $\vert\Psi(t)\vert^2$; b) the average displacements $\beta_{n}(\mathring{A})$; c) the electron current $i(t)$ (expressed in $\mu$biot); d) the frequency spectrum $\vert\tilde{i}(\nu)\vert^2$ along the sequence of $N=66$ nucleotides  with the initial conditions $E_0=0.79$ eV ($E'=120$) and the initial electron excitation peaked around the site $n_0=N/3$.  The other parameters are the same of Fig. \ref{amplitude3452}.}
 \label{amplitude3426}
 \end{figure}

\begin{figure}
\includegraphics[width=0.45\columnwidth]{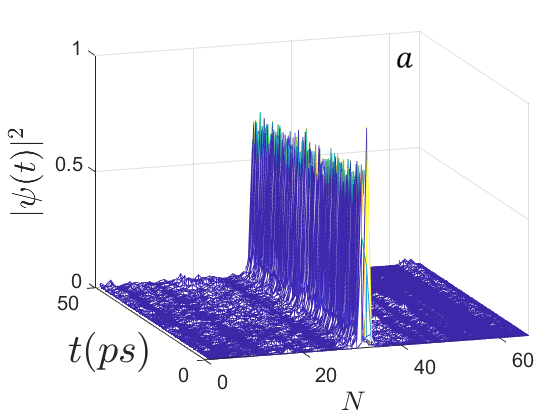}
\includegraphics[width=0.45\columnwidth]{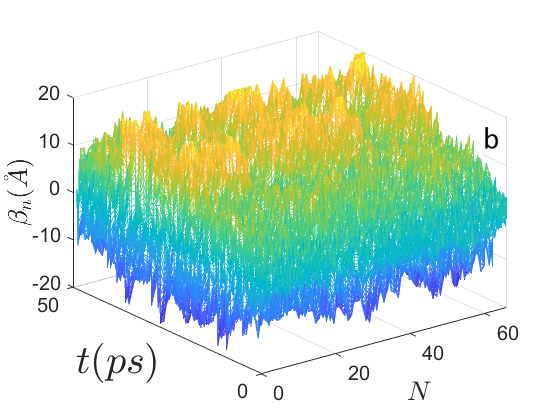}
\includegraphics[width=0.45\columnwidth]{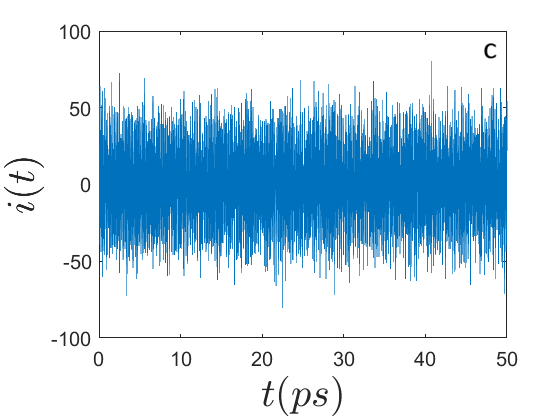}\hskip 1.5truecm
\includegraphics[width=0.45\columnwidth]{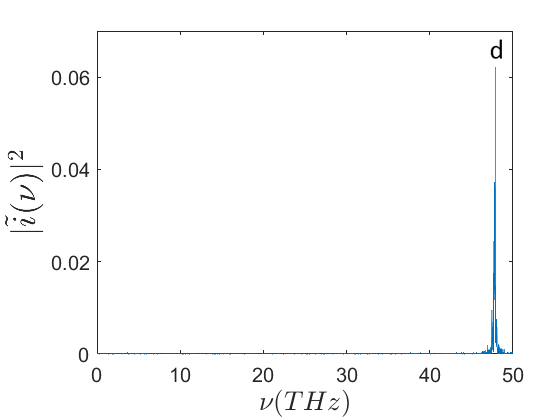}      
   \caption{Time evolution of: a) the electron probability amplitude $\vert\Psi(t)\vert^2$; b) the average displacements $\beta_{n}(\mathring{A})$; c) the electron current $i(t)$ (expressed in $\mu$biot); d) the frequency spectrum $\vert\tilde{i}(\nu)\vert^2$ along the sequence of $N=66$ nucleotides  with the initial conditions $E_0=0.79$ eV ($E'=120$) and the initial electron excitation peaked around the site $n_0=N/2$.  The other parameters are the same of Fig. \ref{amplitude3452}.}
 \label{amplitude3421}
 \end{figure}

\begin{figure}
\includegraphics[width=0.45\columnwidth]{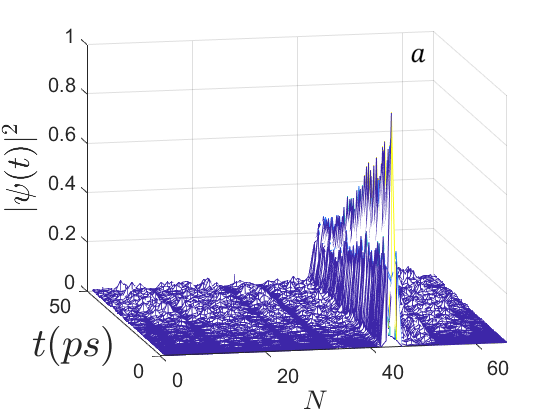}
\includegraphics[width=0.45\columnwidth]{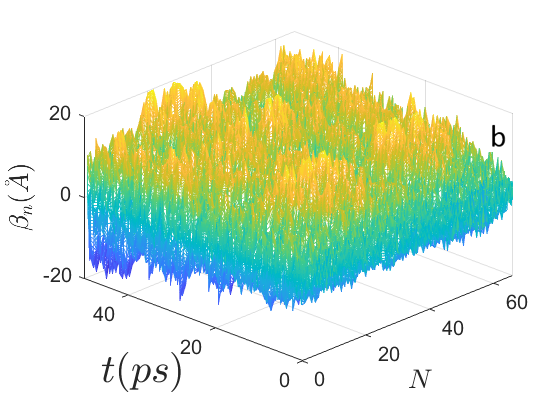}
\includegraphics[width=0.45\columnwidth]{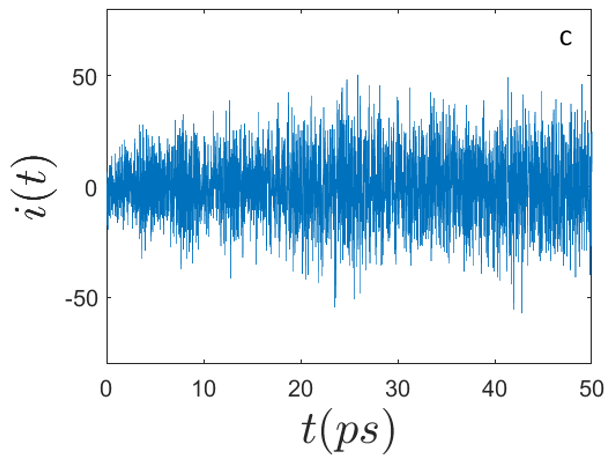}\hskip 1.5truecm
\includegraphics[width=0.45\columnwidth]{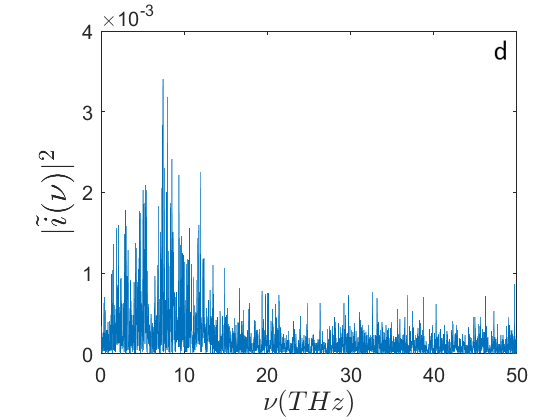}      
   \caption{Time evolution of: a) the electron probability amplitude $\vert\Psi(t)\vert^2$; b) the average displacements $\beta_{n}(\mathring{A})$; c) the electron current $i(t)$ (expressed in $\mu$biot); d) the frequency spectrum $\vert\tilde{i}(\nu)\vert^2$ along the sequence of $N=66$ nucleotides with the initial conditions $E_0=0.9$ eV ($E'=136.7$) and the initial electron excitation peaked around the site $n_0=2N/3$.  The other parameters are the same of Fig. \ref{amplitude3452}.}
 \label{amplitude3432}
 \end{figure}

 \begin{figure}[h!]
\includegraphics[width=0.45\columnwidth]{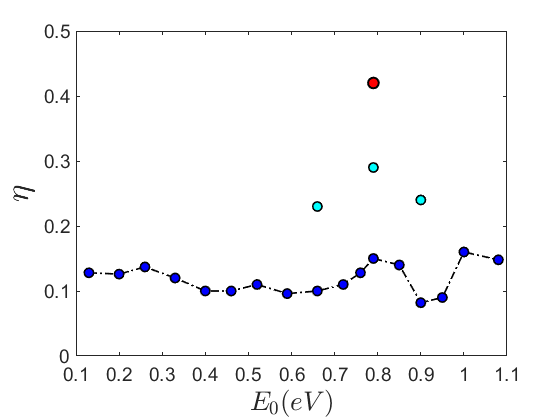}
\includegraphics[width=0.45\columnwidth]{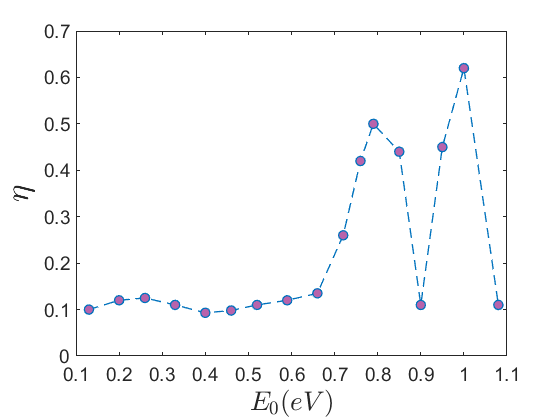}
   \caption{The normalized spectral entropy $\eta$ versus the initial energy $E_0$ for a DNA protein. The parameters are the same of those reported for Fig. \ref{amplitude3452}. The left panel refers to  $N=398$ bp where the dark blue circles correspond to the initial electron excitation peaked around the site  $n_0=N/2$,
   the light blue circles correspond to  $n_0=N/3$, and the red circle correspond to $n_0=2N/3$. The right panel refers to $N=66$ bp and the initial electron excitation is peaked around the site  $n_0=N/2$. }
          \label{entropy}
   \end{figure}

\clearpage
\section*{Appendix A}
\noindent CGGCTGTCATCACTTAGACCTCACCCTGTGGAGCCACACCCTAGGGTT\\
GGCCAATCTACTCCCAGGAGCAGGG AGGGCAGGAGCCAGGGCTGGGCATAAA\\
AGTCAGGGCAGAGCCATCTATTGCTTACATTTGCTTCTGACACAACTGTGTTCACT\\
AGCAACCTCAAACAGACACCATGGTGCATCTGACTCCTGTGGAGAAGTCTGCCGT\\
TACTGCCCTGTGGGGCAAGGTGAACGTGGATGAAGTTGGTGGTGAGGCCCTGGG\\
CAGGTTGGTATCAAGGTTACAAGACAGGTTTAAGGAGACCAATAGAAACTGGGCAT\\ GTGGAGACAGAGAAGACTCTTGGGTTTCTGATAGGCACTGACTCTCTCTGCCTATT\\
GGTCTATTTTCCCACCCTTAG
\\

\noindent Gene sequence (398 bp) coding for one of the $\beta$ subunits of the haemoglobin molecule (HBB)

\section*{Appendix B}
\noindent ATGGCTAATGACCGAGAATAGGGATCCGAATTCAATATTGGTACCTA\\
CGGGCTTTGCGCTCGTATC


\noindent Sequence of an oligonucleotide (66 bp) containing a cleavage subsequence of the EcoRI enzyme


\appendix

\end{document}